\documentstyle[11pt]{article}
\input {epsf}
\def\Lm{\Lambda}
\newcommand{\beq}{\begin{equation}}
\newcommand{\eeq}{\end{equation}}
\newcommand{\beqy}{\begin{eqnarray}}
\newcommand{\eeqy}{\end{eqnarray}}

\def\cW{{\cal W}}

\def\indspace{\hspace*{1.0em} }
\catcode`\@=11
\def\numberbysection{\@addtoreset{equation}{section}
\def\theequation{\arabic{section}.\arabic{equation}}}
\def\appendix{\setcounter{section}{0}
        \def\thesection{Appendix \Alph{section}}
        \def\theequation{\Alph{section}.\arabic{equation}}}
\begin{document}
\pagestyle{empty}
%\begin{flushleft}
%{\it    Tohoku Institute of Technology\\
%	Kanazawa Gakuin University}
%\end{flushleft}
\begin{flushright}
 hep-th/0203209\\
 March 2002
\end{flushright}
\renewcommand{\thefootnote}{\fnsymbol{footnote}}
%\vspace{0.5in}
\vfill
\begin{center}\Large{\bf 
  Geometrical Construction of Heterogeneous\\ Loop Amplitudes
  in 2D Gravity
}\\
\vspace{1cm}
\normalsize\ Masahiro ANAZAWA \footnote[2]
{E-mail address: anazawa@tohtech.ac.jp} 
\hspace{2mm}
and
\hspace{4mm} 
Atushi ISHIKAWA \footnote[3]{
E-mail address: ishikawa@kanazawa-gu.ac.jp}\\
\vspace{10mm}
%$~^{1}$
$~^{\dagger}$
	{\it Tohoku Institute of Technology,
	Sendai 982-8577, Japan\\}
\vspace{5mm}
%and\\
%\vspace{5mm}
%$~^{2}$
$~^{\ddagger}$ 
	{\it Kanazawa Gakuin University,
	 Kanazawa 920-1392, Japan \\}
\vspace{10mm}
%(Received August 22, 2000)
\vspace{0.1in}
\end{center}
\renewcommand{\thefootnote}{\arabic{footnote}}
\setcounter{footnote}{0}
\vspace{10mm}
%\vspace{0.4in}
\begin{abstract}
\baselineskip 17pt
 We study a disk amplitude
 which has a complicated heterogeneous matter configuration on the boundary
 in a system of the $(3,4)$ conformal matter coupled to 
 two-dimensional gravity.
 It is analyzed using the two-matrix chain model in the large $N$ limit.
 We show that the disk amplitude calculated by Schwinger-Dyson equations
 can completely be reproduced  through purely geometrical consideration.
 From this result, we speculate that all heterogeneous loop amplitudes
 can be derived from the geometrical consideration and the consistency
 among relevant amplitudes.
\vspace{5mm}
%\begin{flushleft}
%PACS nos.: 04.60.Nc, 11.25.Pm\\
%Keyword: Matrix Model; Two-dimensional Gravity
%\end{flushleft}
\end{abstract}
\vfill

\newpage
\baselineskip 17pt
\pagestyle{plain}
\setcounter{page}{1}
%%%%%%%%%%%%%%%%%%%%%%%%%%%%%%%%%%%%%%%%%%%%%%%%%%%%%%%%%%%%%%%%%%%%%%%%%%%%%%%
%%%%%%%%%%%%%%%%%%%%%%%% Introduction %%%%%%%%%%%%%%%%%%%%%%%%%%%%%%%%%%%%%%%%%
%%%%%%%%%%%%%%%%%%%%%%%%%%%%%%%%%%%%%%%%%%%%%%%%%%%%%%%%%%%%%%%%%%%%%%%%%%%%%%%
\numberbysection
\section{Introduction}
\label{introduction}

\indspace
The $(m,m+1)$ unitary minimal conformal model coupled to 
two-dimensional gravity can be
analyzed by the two-matrix chain model 
 or the multi-matrix chain model \cite{HI}-\cite{DKK}
 under appropriate scaling limits.
% \cite{GM}.
In two-dimensional gravity,
loop amplitude is  one of the most important
physical quantities, which is the amplitude for a surface with
loop-like boundaries.
The $(m,m+1)$ minimal model has $m-1$ matter states \cite{Pasq}, so
in general, we have to deal with loops on which matter configurations
 are not restricted to homogeneous ones 
 \cite{GN}-\cite{AI00}.  

In Refs.~\cite{AIT2, AI00}, some disk amplitudes whose boundaries have
 heterogeneous matter states  are studied in the case of
 the tricritical Ising model ($m=4$ case).
 It is found that the obtained amplitudes have simple geometrical
 interpretation.
 The heterogeneous loops make interesting splitting and
 sticking interaction regularly.
 Here a question arises. Can we decide any heterogeneous loop amplitudes
 only from the geometrical consideration?
 In other words, does the geometrical consideration give enough
 information to decide any heterogeneous loop  amplitudes?
 
 In order to address this question, in this paper, we study a disk
 amplitude with rather complicated boundary condition
 in the case of the Ising model ($m=3$ case).
 We show that the amplitude calculated by Schwinger-Dyson equations
 can be reproduced through purely geometrical consideration.
 We speculate that any heterogeneous loop
 amplitudes can be decided geometrically.
\\

%%%%%%%%%%%%%%%%%%%%%%%%%%%%%%%%%%%%%%%%%%%%%%%%%%%%%%%%%%%%%%%%%%%%%
%%%%%%%%%%%%%%%%%%%%%%%%%%%%%%%%%%%%%%%%%%%%%%%%%%%%%%%%%%%%%%%%%%%%%
\section{Amplitude from Schwinger-Dyson equations}
\label{fromSD}

\indspace
In this section, we examine a disk amplitude with heterogeneous matter
boundary condition using Schwinger-Dyson equations. 
Let us start with the action of the two-matrix model
\beq
S(A,B) = \frac{N}{\Lambda} 
                   {\rm tr} \left\{ U(A) + U(B)  - A B  \right\}.
\label{2-matrix model Action}
\eeq
Here $A$ and $B$ are $N \times N$ unitary matrix variables,
and $\Lm$ is the bare cosmological constant.
As a critical potential which realizes the Ising model coupled
to two-dimensional gravity, we take
\beq
U(\phi)  =  \alpha\, \phi + \frac{\beta}{2} \phi^2 + \frac{\gamma}{3} \phi^3 \;,
\quad
(\alpha,\beta,\gamma)=(3,-3,-1) \;. 
\eeq 
This can be found using 
the orthogonal polynomial method \cite{DKK, AIT2}.

In this paper we will focus on the amplitude
\beq
W_{ABAB}(p_1, q_1, p_2, q_2)=
\sum_{k,l,m,n=0}^{\infty}  
\frac{\Lm}{N} \left\langle {\rm tr}(A^k B^l A^m B^n) \right\rangle
 p_1^{-k-1} q_1^{-l-1} p_2^{-m-1} q_2^{-n-1}
\eeq 
and its continuum universal counterpart 
$w_{ABAB}(\zeta_{A1},\zeta_{B1},\zeta_{A2}, \zeta_{B2}, t)$ 
in the large $N$ limit.
 Here $p_i$ and $q_i$ are bare boundary cosmological constants,
 $\zeta_{A_i}$ and  $\zeta_{B_i}$  are their renormalized
 counterparts, and $t$ is the renormalized cosmological constant.
The insertion of the operator ${\rm tr}(A^k B^l A^m B^n)$ makes
a loop-like boundary on which a part with spins up and 
a part with spins down appear
by turns. In the large $N$ limit, the corresponding continuum amplitude becomes
 the disk amplitude whose boundary consists of four arcs with
 distinct matter states.

%In this section we will derive the disk amplitude using Schwinger-Dyson equations. 
Let us start with the Schwinger-Dyson equation
\beq
0 = \sum_a \int dAdB \frac{\partial}{\partial A^a} 
        \left\{
              {\rm tr} \left( A^{k} t^a B^{l} A^m B^n \right) e^{-S(A,B)} 
        \right\} \;,
\label{SD_0}
\eeq
here the hermitian matrix $A$ is expressed
as $A=\sum_a A^a t^a$ using the bases of hermitian matrices
 $\left\{ t^a \right\}$.
Rewriting Eq.~(\ref{SD_0}) explicitly, we obtain
\beqy
0 &=& \sum_{i=0}^{k-1}[A^i][A^{k-i-1}B^l A^m B^n]
 +\sum_{i=0}^{m-1}[B^l A^{m-i-1}][A^k A^i B^n]
 -\alpha [A^k B^l A^m B^n] \nonumber \\
 &&-\beta [A^{k+1} B^l A^m B^n]
 -\gamma [A^{k+2} B^l A^m B^n] +[A^k B^{l+1} A^m B^n]
 \;,
\label{SD_1}
\eeqy
here we use the notation
$[A^k B^l \cdots]=\frac{\Lm}{N}\langle {\rm tr}(A^k B^l \cdots) \rangle $ .
Multiplying $p_1^{-k-1} q_1^{-l-1} p_2^{-m-1} q_2^{-n-1}$ and
summing up with respect to $k$, $l$, $m$ and $n$ indices, we obtain 
the equation in terms of resolvent amplitudes
\beqy
0 &=& \left( W_A(p_1)-v(p_1)+q_1 \right)\, W_{ABAB}(p_1,q_1,p_2,q_2)
 +\left( W_{BA}(q_1,p_2)-1 \right)\, W_{AAB}(p_1,p_2,q_2)
 \nonumber \\
 &&+(\beta+\gamma p_1) W_{BAB}(q_1,p_2,q_2) 
 +\gamma W^{(A)}{}_{BAB}(q_1,p_2,q_2) \;.
\label{SD_ABAB}
\eeqy
In Eq.~(\ref{SD_ABAB}) and in the following, we  use the notations
%\beq
 $v(p)=\alpha +\beta p +\gamma p^2 \;$,
%\eeq
\beq
W_{M_1 M_2 \,\cdots\,}(r_1,r_2,\,\cdots\,)
 =\sum_{k,l,\,\cdots\,=0}^{\infty} \frac{\Lm}{N} 
 \left\langle {\rm tr}(M_1^{k} M_2^{l} \,\cdots\,) \right\rangle
  r_1^{-k-1} r_2^{-l-1} \,\cdots\; ,
\eeq
\beq
W^{(M)}{}_{M_1}{}^{(M')}{}_{M_2 \,\cdots\,}
(r_1,r_2,\,\cdots\,)
 =\sum_{k,l,\,\cdots\,=0}^{\infty} \frac{\Lm}{N} 
 \left\langle {\rm tr}(M M_1^{k} M' M_2^{l} \,\cdots\,) \right\rangle
  r_1^{-k-1} r_2^{-l-1} \,\cdots\; ,
\eeq
where $M$, $M'$ and $M_i$ denote the matrices $A$ or $B$.
Using the relations
$ W_{BA}(q_1,p_2) = W_{AB}(p_2,q_1) $ and
$ W_{BAB}(q_1,p_2,q_2) = W_{AAB}(q_2,q_1,p_2) $,
%$ W_{BAB}(q_1,p_2,q_2) = W_{BBA}(q_2,q_1,p_2) $ ,
we observe that
$W_{ABAB}$ can be expressed in terms of $W_A$, $W_{AB}$, $W_{AAB}$
%, $W_{BBA}$
and $W^{(A)}{}_{BAB}$.

As for $W^{(A)}{}_{BAB}$, 
from the resolvent expression for the $l=0$ case of Eq.~(\ref{SD_1})
and the relations
$W^{(A)}{}_{AB}(p_2,q_2) =p_2 W_{AB}(p_2,q_2) -W_B(q_2)$
and
$W_A{}^{(B)}{}_{AB}(p_1,p_2,q_2)=W^{(A)}{}_{BAB}(p_2,q_2,p_1)$,
we obtain
\beqy
W^{(A)}{}_{BAB}(p_2,q_2,p_1)
&=&
 -\left( W_A(p_1)+W_A(p_2)-v(p_1) \right)\, W_{AAB}(p_1,p_2,q_2)
\nonumber \\
&&
 -\left( \beta+\gamma (p_1+p_2)\right) W_{AB}(p_2,q_2) 
 +\gamma W_B(q_2) \;.
\label{SD_(A)BAB}
\eeqy
Combining Eqs.~(\ref{SD_ABAB}) and (\ref{SD_(A)BAB}), we find that 
$W_{ABAB}$ can be expressed in terms of 
$W_A$, $W_B$, $W_{AB}$
and $W_{AAB}$.
Furthermore using the relations 
\beqy
&&W_{AAB}(p_1,p_2,q) 
 =-\frac{ W_{AB}(p_1,q)-W_{AB}(p_2,q) }{p_1-p_2} \,, 
 \\
 \label{W_AAB}
&&W_{AB}(p,q)
 = \frac{ W_A(p)-(\beta+\gamma p)W_B(q) -\gamma {W_B}^{(A)}(q) }
        { W_A(p)-v(p)+q } \,,
 \\
&&W_B{}^{(A)}(q)
= -\left( W_B(q)-v(q) \right) W_B(q)
  -(\beta+\gamma q)\Lm -\gamma [A] \,,
\eeqy
we find, in the end, that $W_{ABAB}$ can be expressed in terms of 
$W_{A}$ and $W_{B}$.
(Here the last two equations are obtained in a similar way
to Refs. \cite{GN,Staudacher,SY,AIT2,AI00}.)

The homogeneous disk amplitudes $W_{A}$ and $W_{B}$ satisfy a third order
equation. They are expanded as
\beq
 W_A(p) =  W_B(p) =
 3 -2 a \zeta +\left( \frac{a}{2} \right)^{4/3}\, w(\zeta)
 + {\cal O}(a^{5/3}) \;,
\label{homo disk}
\eeq
here $a$ is the lattice spacing and
%the continuum homogeneous disk amplitude \cite{MSS}
%can be identified as
\beq
 w(\zeta) =\Bigl(\, \zeta +\sqrt{\zeta^2 -t} \;\;\Bigr)^{4/3}
            +\Bigl(\, \zeta -\sqrt{\zeta^2 -t} \;\;\Bigr)^{4/3}
\eeq
is the continuum homogeneous disk amplitude \cite{MSS}
under the renormalization 
$\Lm=10-a^2 t$ and $p=a \zeta$. 
This amplitude satisfies the third order 
equation
\beq
 w(\zeta)^3 -3 t^{4/3} w(\zeta) -16 \zeta^4 +16t \zeta^2
 -2 t^2 =0 \;.
\label{identity} 
\eeq
Combining all of the above equations, we obtain the continuum heterogeneous
disk amplitude 
%$w_{ABAB}(\zeta_{A_1},\zeta_{B_1},\zeta_{A_2},\zeta_{B_2})$.
$w_{ABAB}$. From the explicit calculation,
we find that terms in  ${\cal O}(a^{5/3})$ in
Eq.~(\ref{homo disk}) do not affect  the calculation of $w_{ABAB}$,
and $w_{ABAB}$ is expressed
only in terms of $w(\zeta_{A_i})$ and $w(\zeta_{B_j})$. From the
$a^1$ order terms of $16~W_{ABAB}$, we obtain
the continuum disk amplitude $w_{ABAB}$ as follows:
\beqy
\lefteqn{ w_{ABAB}(\zeta_{A_1},\zeta_{B_1},\zeta_{A_2},\zeta_{B_2}) }
\nonumber \\
&=&
-\frac{w(\zeta_{A_1}) + w(\zeta_{A_2})}{ \zeta_{A_1}-\zeta_{A_2} }
\left( 
 \frac{w(\zeta_{A_1})^2}{(\zeta_{A_1}+\zeta_{B_1})(\zeta_{A_1}+\zeta_{B_2})}
-\frac{w(\zeta_{A_2})^2}{(\zeta_{A_2}+\zeta_{B_1})(\zeta_{A_2}+\zeta_{B_2})}
\right)
\nonumber \\
&&
-\frac{w(\zeta_{B_1}) + w(\zeta_{B_2})}{ \zeta_{A_1}-\zeta_{A_2} }
\left( 
 \frac{w(\zeta_{A_1})^2}{(\zeta_{A_1}+\zeta_{B_1})(\zeta_{A_1}+\zeta_{B_2})}
-\frac{w(\zeta_{A_2})^2}{(\zeta_{A_2}+\zeta_{B_1})(\zeta_{A_2}+\zeta_{B_2})}
\right)
\nonumber \\
&&
-\frac{w(\zeta_{B_1})\, w(\zeta_{B_2})}{ \zeta_{A_1}-\zeta_{A_2} }
\left( 
 \frac{w(\zeta_{A_1})}{(\zeta_{A_1}+\zeta_{B_1})(\zeta_{A_1}+\zeta_{B_2})}
-\frac{w(\zeta_{A_2})}{(\zeta_{A_2}+\zeta_{B_1})(\zeta_{A_2}+\zeta_{B_2})}
\right)
\nonumber \\
&&
-3t^{4/3} \left( w(\zeta_{A_1})+w(\zeta_{A_2}) \right)
 \frac{ \zeta_{A_1}+\zeta_{A_2}+\zeta_{B_1}+\zeta_{B_2} }
 {(\zeta_{A_1}+\zeta_{B_1})(\zeta_{A_2}+\zeta_{B_1})
 (\zeta_{A_1}+\zeta_{B_2})(\zeta_{A_2}+\zeta_{B_2})}
\nonumber \\
&&
+ \left( A_1 \leftrightarrow B_1, \; A_2 \leftrightarrow B_2 \right).
\label{w_ABAB_fromSD}
\eeqy
Here the last expression 
$\left( A_1 \leftrightarrow B_1, \; A_2 \leftrightarrow B_2 \right)$
denotes the terms which are obtained by interchanging $A_i$ and $B_i$
in all of the preceding terms. 
As the continuum part, we have taken terms which are non-analytic
in all of the variables $\zeta_{A_i}$, $\zeta_{B_j}$ and $t$.
We should notice that the form of the amplitude (\ref{w_ABAB_fromSD}) 
is not unique because of 
the identity equation (\ref{identity}).
This uncertainty, however, does not arise in the geometrical consideration
in the next section.

%%%%%%%%%%%%%%%%%%%%%%%%%%%%%%%%%%%%%%%%%%%%%%%%%%%%%%%%%%%%%%%%%%
%%%%%%%% section 3 %%%%%%%%%%%%%%%%%%%%%%%%%%%%%%%%%%%%%%%%%%%%%%%
\section{Amplitude from geometrical consideration}
\label{fromGC}

\indspace
In this section, we will show that the heterogeneous disk
amplitude $w_{ABAB}$ obtained in section 2 can be
reproduced through purely geometrical consideration.

\begin{figure}[htb]
\begin{center}
 \begin{minipage}[htb]{0.35\textwidth}
   \epsfxsize = 1.0\textwidth
   \epsfbox{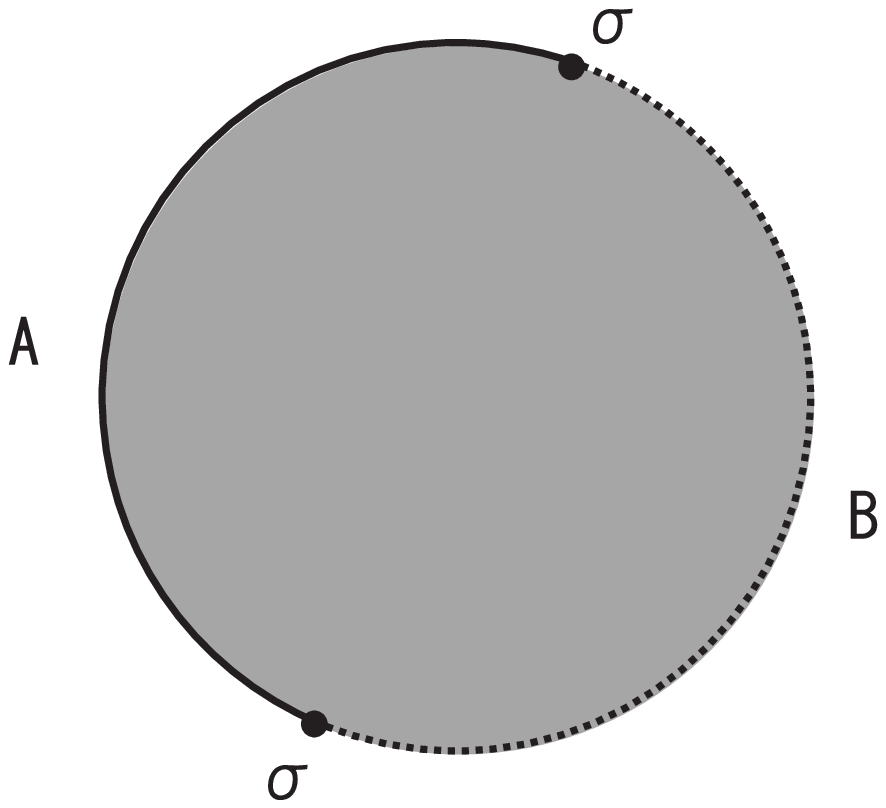}
%   \caption{a1}
%   \label{}
 \end{minipage}
 \caption{
   The two spin operators $\sigma$
   make discontinuities in the matter configuration
   on the boundary.
 }
 \label{p0}
\end{center} 
\end{figure}

\begin{figure}[htb]
\begin{center}
 \begin{minipage}[htb]{0.3\textwidth}
   \epsfxsize = 1.0\textwidth
   \epsfbox{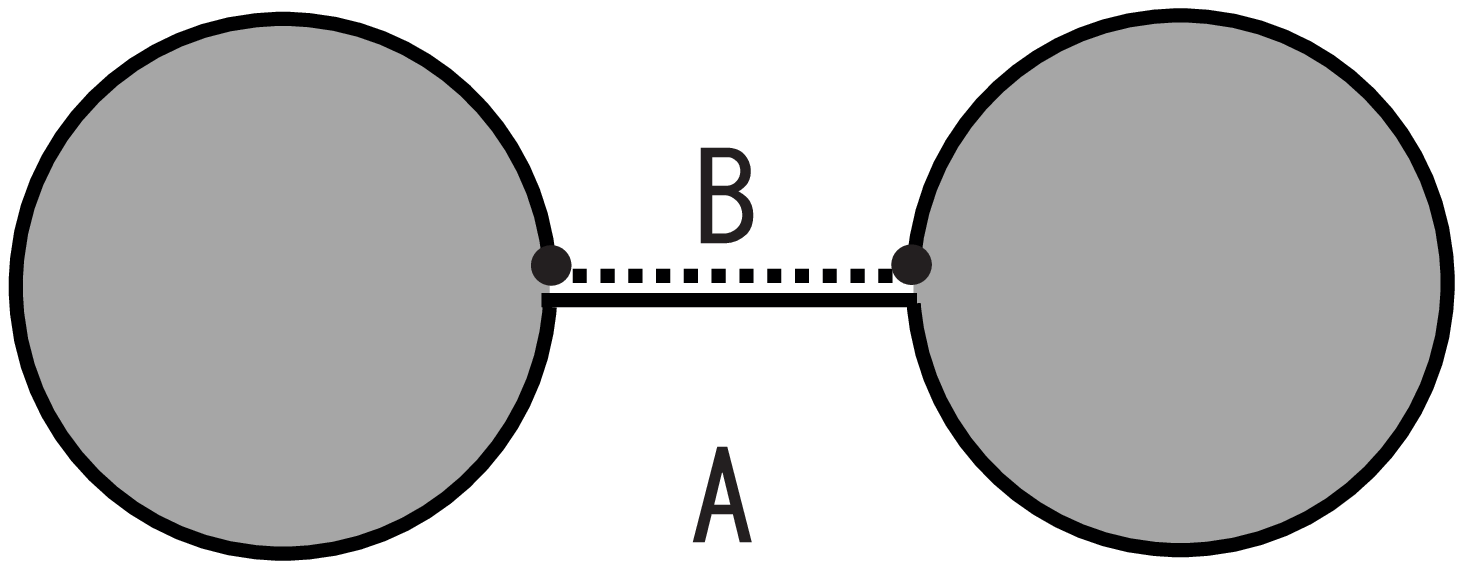}
%   \caption{}
%   \label{}
 \end{minipage}
% \hfill
 \hspace{0.4cm}
 \begin{minipage}[htb]{0.3\textwidth}
   \epsfxsize = 1.0\textwidth
   \epsfbox{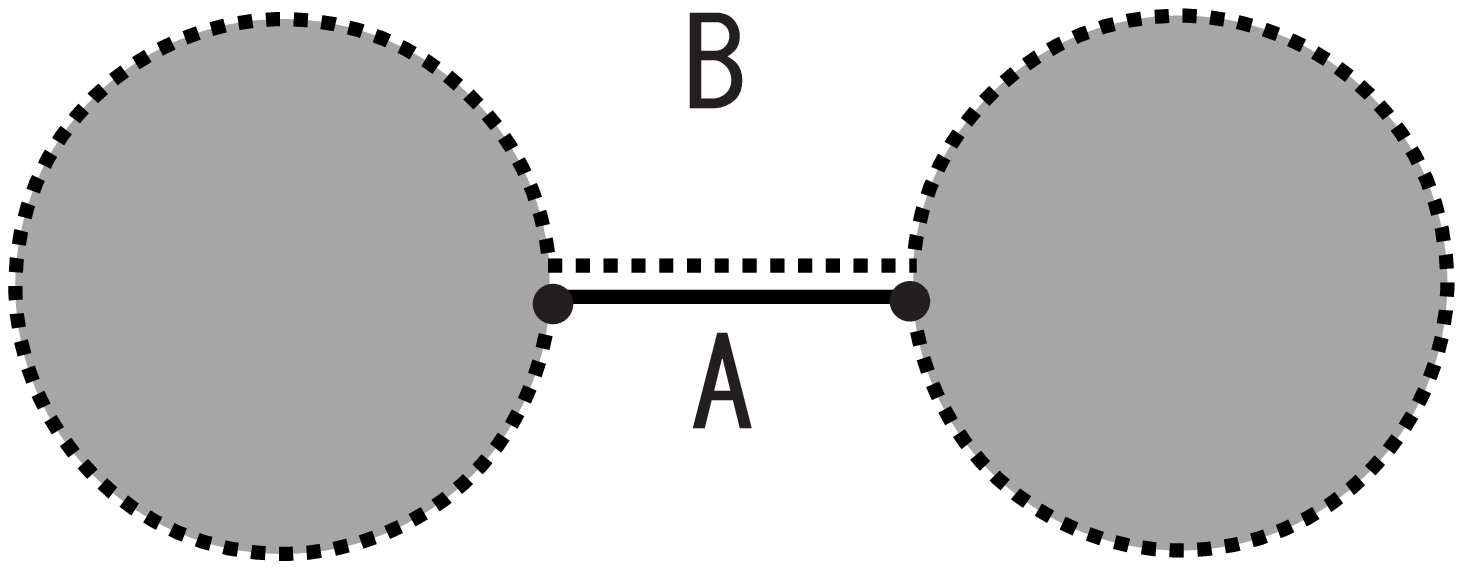}
%  \caption{}
%  \label{}
 \end{minipage}
  \hspace{0.4cm}
 \begin{minipage}[htb]{0.3\textwidth}
   \epsfxsize = 1.0\textwidth
   \epsfbox{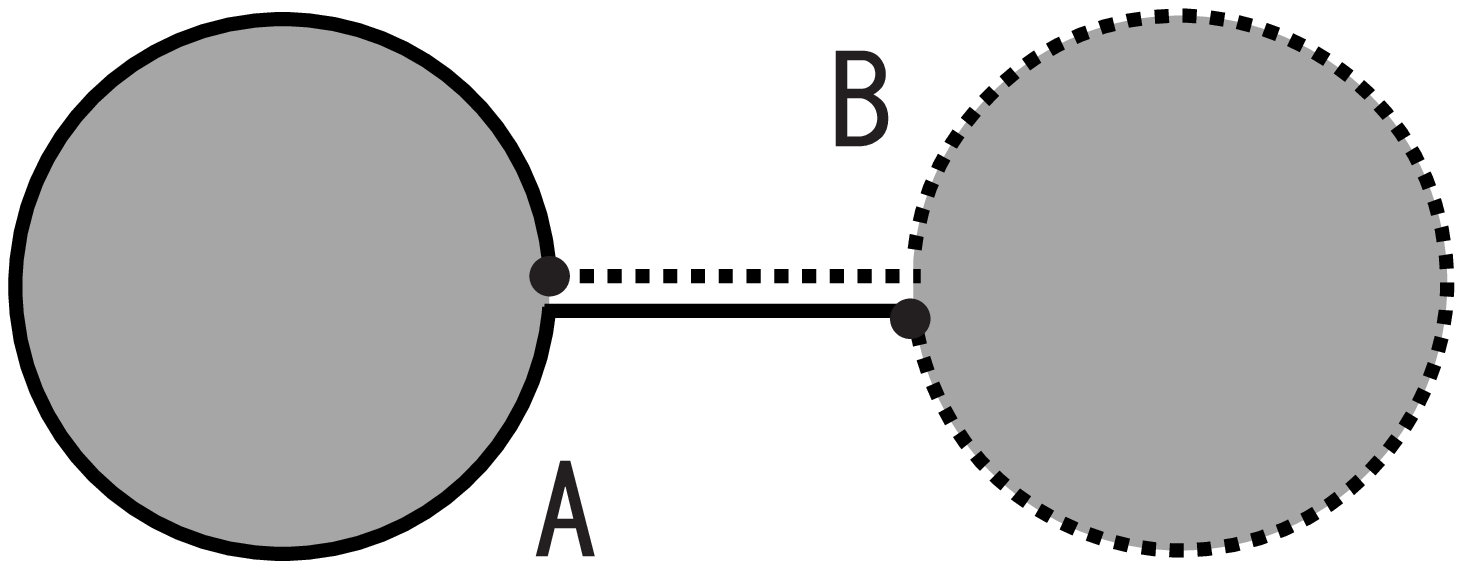}
%  \caption{}
%  \label{}
 \end{minipage}
 \vspace{5mm}
 \caption{
    The diagrams representing the first three terms
    in Eq.~(\ref{cW_AB})
  }
 \label{AB}
\end{center} 
\end{figure}

At first let us briefly review the mechanism of the interaction of
heterogeneous boundary in the case of 
$w_{AB}$ \cite{AIT2}, which is the continuum counterpart of $W_{AB}$.
In this case, the boundary loop consists of two parts
on which the spin state is up or down.
According to boundary conformal field theory,
discontinuity in the matter boundary conditions corresponds to
the insertion of a boundary operator in Refs.~\cite{Cardy,IIKMNS}.
So in the case of $w_{AB}$, we consider that two spin operators
 $\phi_{+-}\equiv \sigma$ are situated at the points where spin state
  changes (see Fig.~\ref{p0}).
The continuum amplitude $w_{AB}$ is calculated to be
 \cite{IIKMNS,SY}
\beq
w_{AB}(\zeta_A,\zeta_B)=
\frac{ w(\zeta_A)^2 +w(\zeta_B)^2 +w(\zeta_A)\,w(\zeta_B)  -3t^{4/3} }
{\zeta_A+\zeta_B} \; .
\label{w_AB}
\eeq
By performing the inverse Laplace transformation,
we obtain the disk amplitude $\cW_{AB}(\ell_A,\ell_B)$ 
in terms of the lengths of the boundaries (see appendix A),
\beqy
\cW_{AB}(\ell_A,\ell_B)&=&{\cal L_A}^{-1} {\cal L_B}^{-1} \left[ 
				w_{AB}(\zeta_A,\zeta_B)	\right]
\nonumber \\
&=&\theta(\ell_A-\ell_B) (\cW*\cW)(\ell_A-\ell_B)
+ \theta(\ell_B-\ell_A) (\cW*\cW)(\ell_B-\ell_A)
\nonumber \\
&&+
\int_0^{{\rm min}(\ell_A,\ell_B)} \!\! d\ell \;
\cW(\ell_A-\ell) \, \cW(\ell_B-\ell)
\;-3t^{4/3} \delta(\ell_A-\ell_B) \; .
\label{cW_AB}
\eeqy From Eq.~(\ref{cW_AB})
we find that the following geometrical interpretation
is possible \cite{AIT2}. 
The original heterogeneous
 loop splits into two homogeneous loops. 
 The first three terms in 
Eq.~(\ref{cW_AB}) correspond to the three diagrams 
in Fig.~\ref{AB} respectively.
This loop splitting phenomenon can be summarized in the following
rules.
\begin{enumerate}
\item
 Boundary A~\footnote{ 
     We refer to the part of boundary on which the spin state is up as 
     boundary A etc, in the same way to \cite{AIT2}.} 
 sticks to boundary B so that 
 a heterogeneous loop splits into homogeneous loops.
 The split disks are connected by the double line
 which is formed by sticking of boundaries A and B.
\item 
 The spin operator $\sigma$ is situated at each end of the double line.  
\end{enumerate}
The spin operator $\sigma$ separates the boundary into two sides.
One side forms homogeneous loop-like boundary and the other 
forms a stuck double line.
We speculate that the spin operator $\sigma$ has the effect to make
discontinuity in the geometrical state of boundary
as well as in the matter state.

Now let us derive $w_{ABAB}$
through a purely geometrical consideration.
For simplicity, we will neglect the terms proportional to
$t^{4/3}$ in Eqs.~(\ref{w_ABAB_fromSD}) and (\ref{w_AB}),
 because they only make the argument complicate.
We can deal with these terms in a similar manner.
The rules of the splitting phenomenon should equally be applied in this case.
We will employ the rules and sum up all possible terms.
Here the weights of the respective terms, however, are not fixed.
The relative weights will be fixed by requiring the consistency with $w_{AB}$
and the expected symmetries of the amplitude $w_{ABAB}$.
The resultant amplitude will completely coincide with that
derived from the Schwinger-Dyson equations
up to overall normalization.

\begin{figure}[htb]
\begin{center}
 \begin{minipage}[htb]{0.35\textwidth}
   \epsfxsize = 1.0\textwidth
   \epsfbox{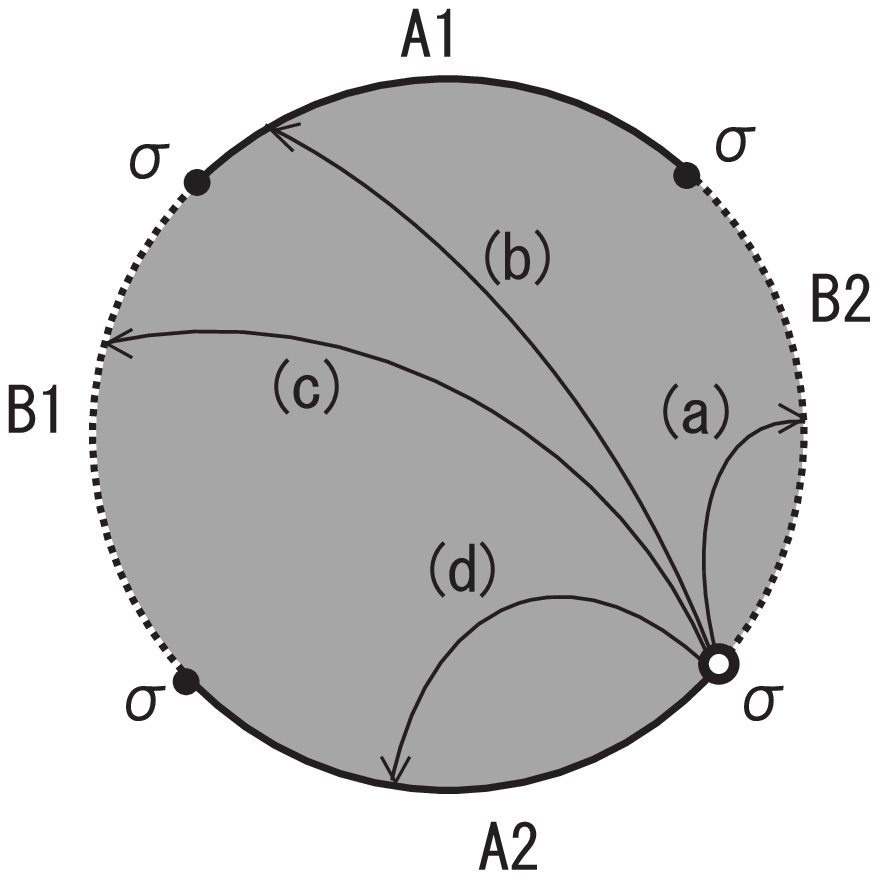}
   \caption{
      There are four cases depending on the behavior of
      the spin operator we focus.
   }
   \label{p1}
 \end{minipage}
 \hspace{20mm}
 \begin{minipage}[htb]{0.35\textwidth}
   \epsfxsize = 1.0\textwidth
   \epsfbox{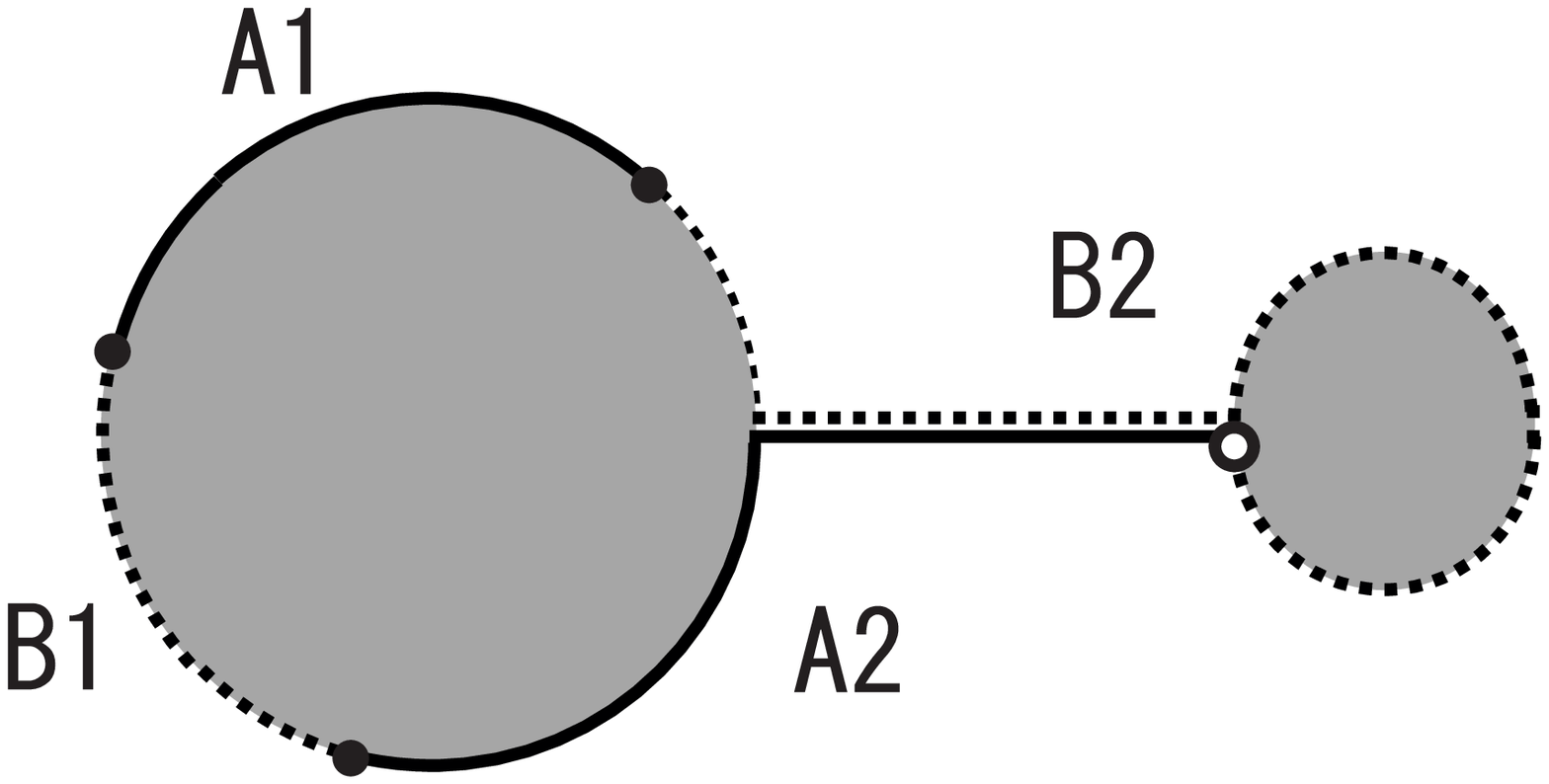}
   \caption{This shows the intermediate situation
   corresponding to the case (a).
   Observe that the spin operator we focus is situated at 
   the right end of the double line.}
   \label{inter}
 \end{minipage} 
% \caption{}
% \label{}
\end{center} 
\end{figure}

\begin{figure}[htb]
\begin{center}
 \begin{minipage}[htb]{0.33\textwidth}
   \epsfxsize = 1.0\textwidth
   \epsfbox{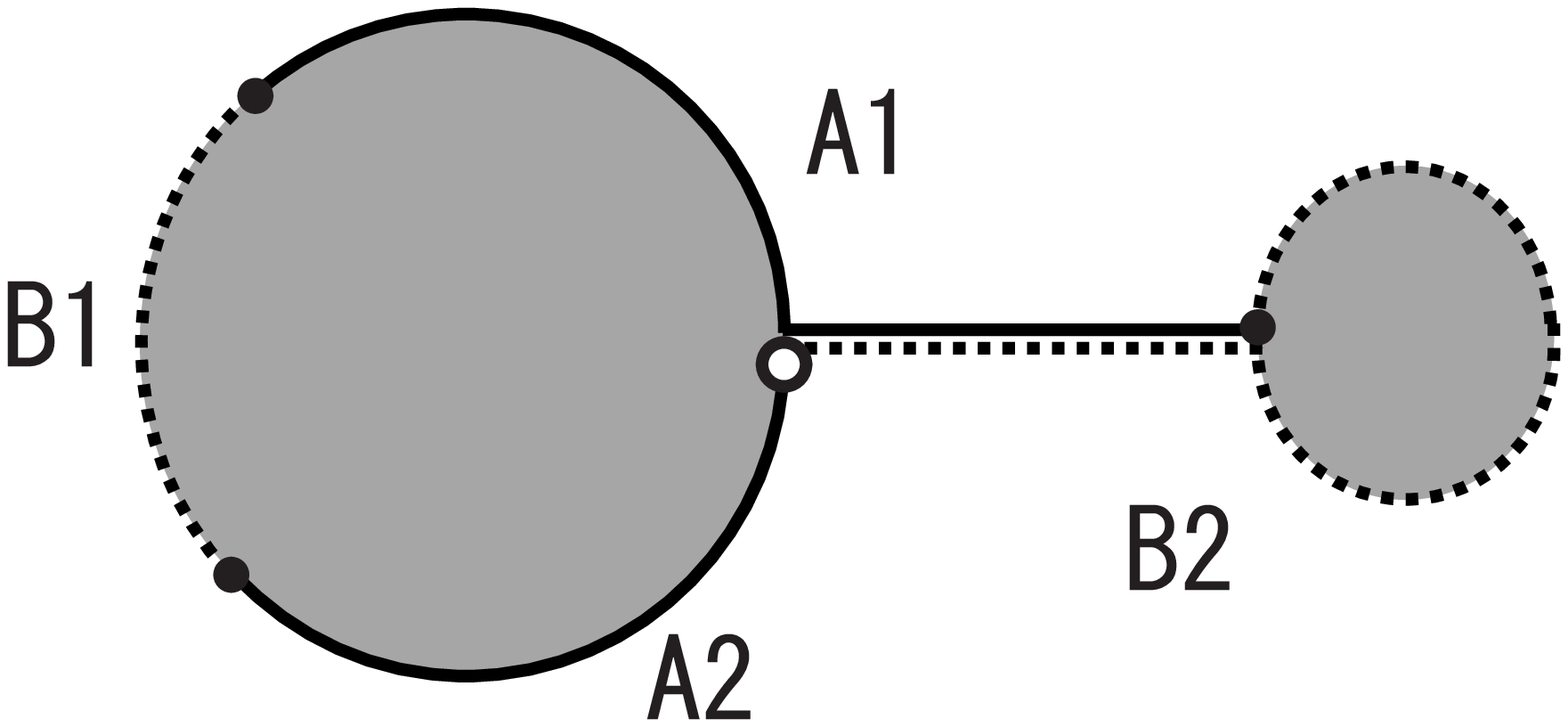}
   \begin{center}\makebox{(b-1)}\end{center}
%   \caption{(b-1)}
%   \label{}
 \end{minipage}
% \hfill
 \hspace{1cm}
 \begin{minipage}[htb]{0.33\textwidth}
%  \mbox{
   \epsfxsize = 1.0\textwidth
   \epsfbox{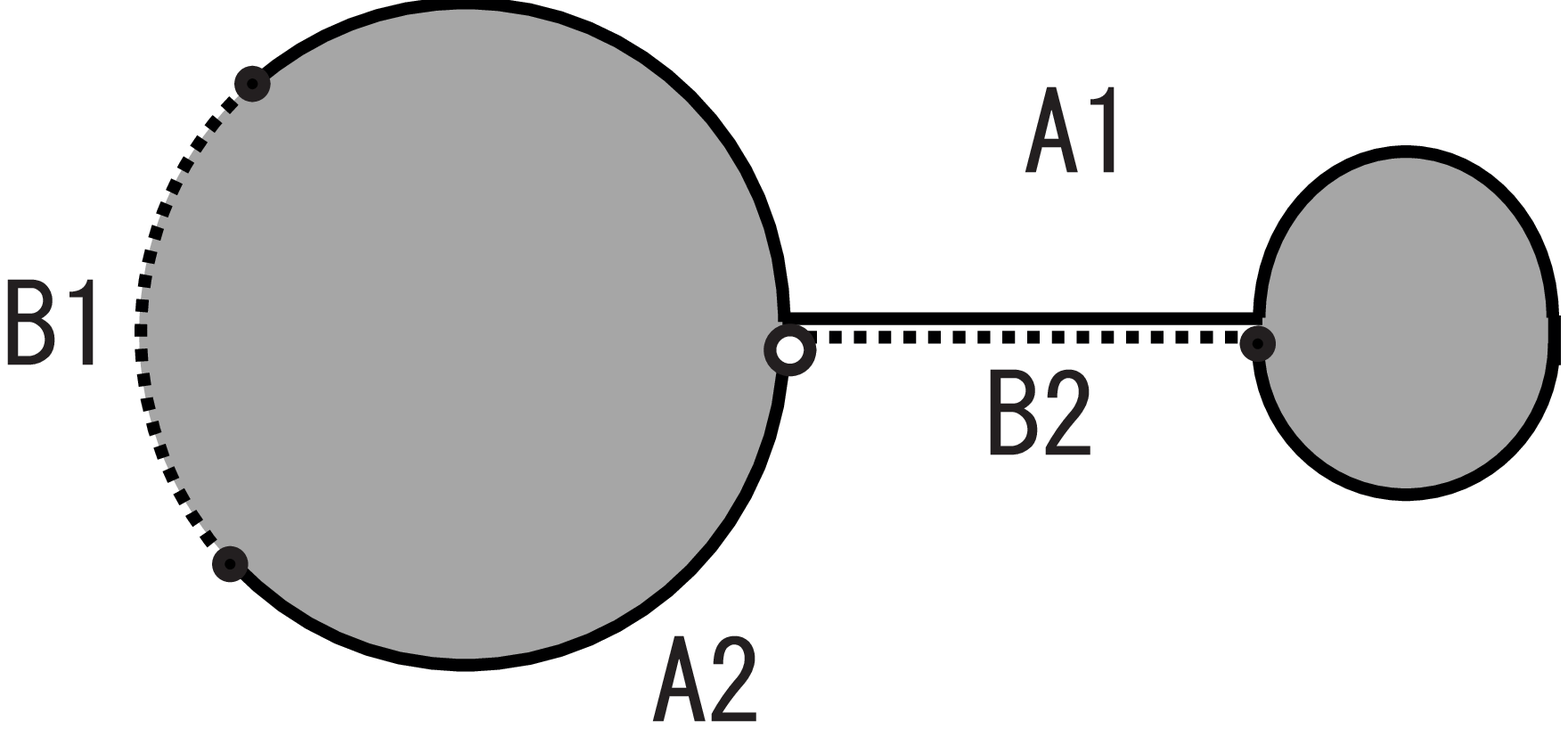}
%   }
   \begin{center}\makebox{(b-2)}\end{center}
%  \caption{(b-2)}
%  \label{}
 \end{minipage}
 \caption{
   The case (b) is also classified into the two sub cases.
 }
 \label{b}
\end{center} 
\end{figure}

\begin{figure}[ht]
\begin{center}
 \begin{minipage}[htb]{0.28\textwidth}
   \epsfxsize = 1.15\textwidth
   \epsfbox{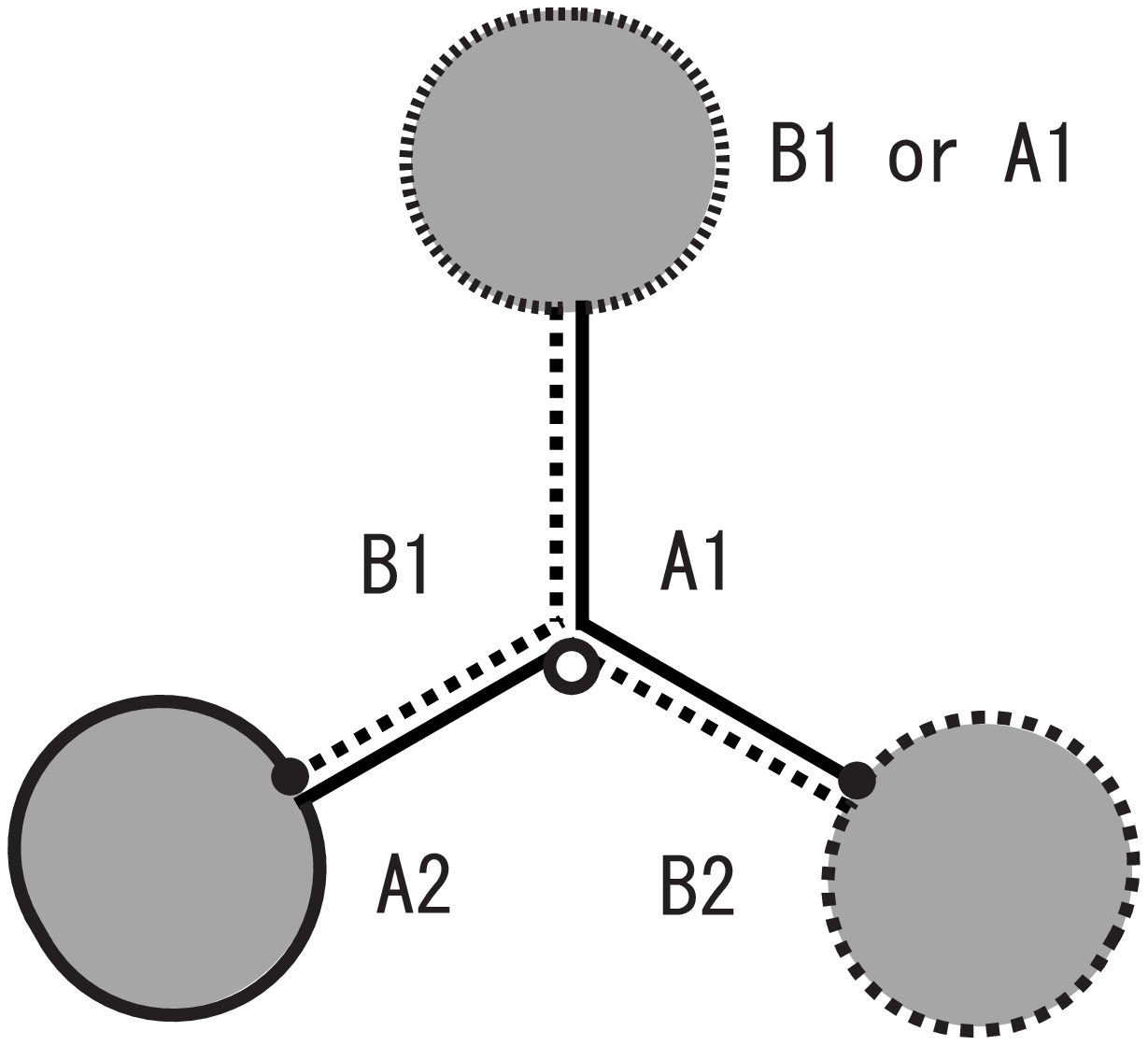}
   \caption{The doubly counted diagram
   between (b-1) and (c-1)}
   \label{e1}
 \end{minipage}
% \hfill
 \hspace{1cm}
 \begin{minipage}[htb]{0.28\textwidth}
   \epsfxsize = 1.0\textwidth
   \epsfbox{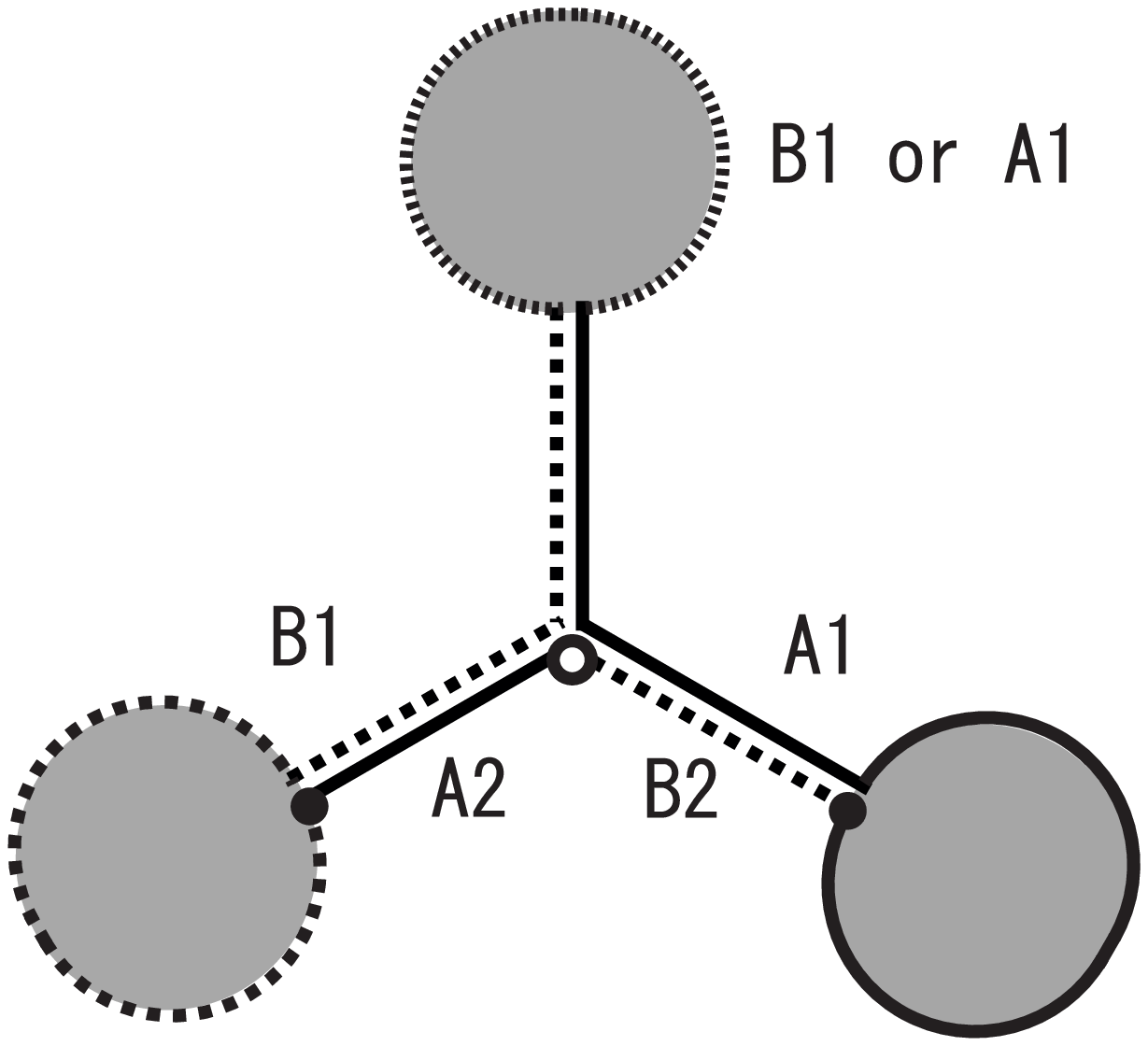}
  \caption{The doubly counted diagram
  between (b-2) and (c-2)}
  \label{e2}
 \end{minipage}
% \caption{}
% \label{}
\end{center} 
\end{figure}

\begin{figure}[ht]
\begin{center}
 \begin{minipage}[htb]{0.28\textwidth}
   \epsfxsize = 1.0\textwidth
   \epsfbox{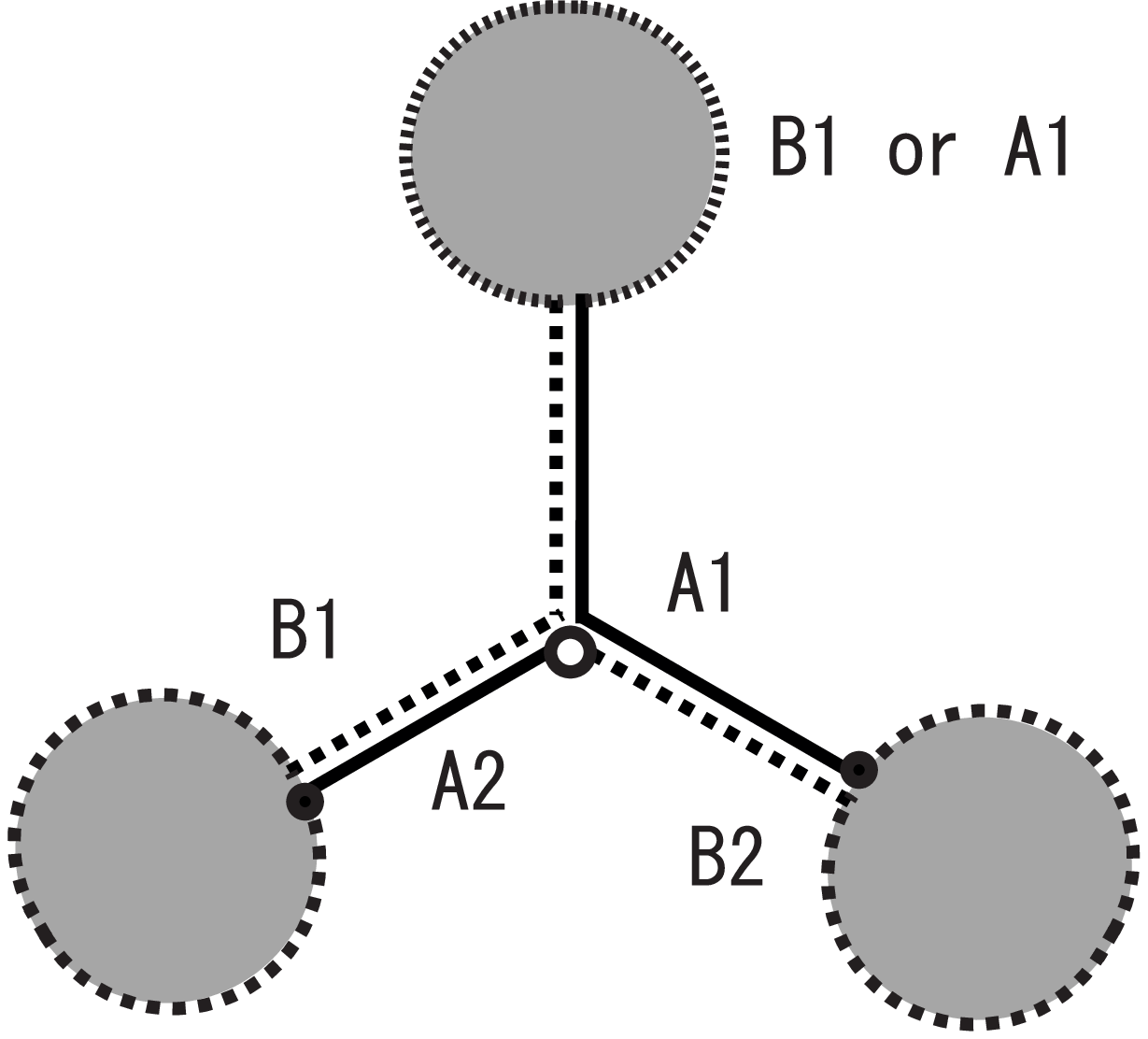}
   \caption{The doubly counted diagram
   between (b-1) and (c-2)}
   \label{e12}
 \end{minipage}
% \hfill
 \hspace{1cm}
 \begin{minipage}[htb]{0.28\textwidth}
   \epsfxsize = 1.0\textwidth
   \epsfbox{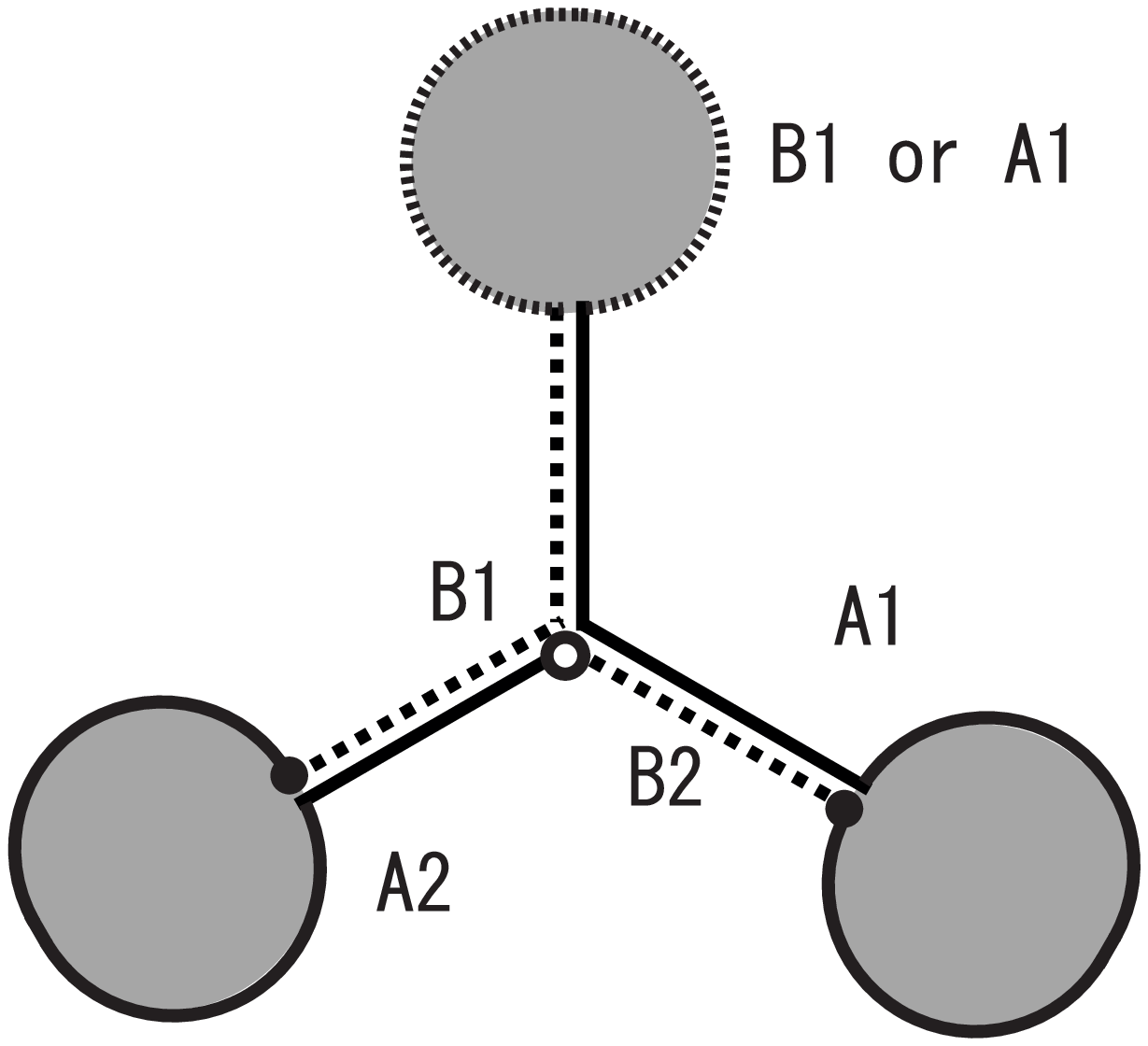}
  \caption{The doubly counted diagram
  between (b-2) and (c-1)}
  \label{e21}
 \end{minipage}
% \caption{}
% \label{}
\end{center} 
\end{figure}

In the following we will explain how to derive the amplitude
$w_{ABAB}$ in detail.
In the case of $w_{ABAB}$, there are four spin operators $\sigma$ on 
the loop-like boundary.
Let us focus on one of them (denoted by $\circ$ in Fig.~\ref{p1}).
After loop splitting, the spin operator we focus is in touch with
another part of the boundary. 
%Depending on with which it is in touch,  
According to the point with which it is in touch,
there are four possibilities,
(a), (b), (c) and (d) in Fig.~\ref{p1}.
As an example let us consider the case (a) concretely.
The spin operator we focus is at one end of the double line (see Fig.~\ref{inter}).
At the other end of the double line, one of the remaining spin 
operators will be situated.
%Depending on which one sits there, 
According to the spin operator situated there,
the case (a) is classified into
three sub cases, (a-1), (a-2) and (a-3) in Fig.~\ref{a}.
Then the amplitude corresponding to the case (a) is written
as follows (see appendix A):
\beqy
\lefteqn{
{\cal G}_a
=
a_1\,
\frac{ w_{AAB}(\zeta_{A_1},\zeta_{A_2},\zeta_{B_1})\, w(\zeta_{B_2 }) }
{\zeta_{A_2}+\zeta_{B_2}}
+a_2\,
\frac{ w_{BBA}(\zeta_{B_1},\zeta_{B_2},\zeta_{A_1})\, w(\zeta_{B_2 }) }
{\zeta_{A_2}+\zeta_{B_2}}
}
 \nonumber \\
&+&\frac{
	\left\{
	a_3\, w(\zeta_{A_1})\, w(\zeta_{A_2})
	+a_4\, w(\zeta_{A_1})\, w(\zeta_{B_1})
	+a_5\, w(\zeta_{B_2})\, w(\zeta_{A_2})
	+a_6\, w(\zeta_{B_2})\, w(\zeta_{B_1})
	\right\} w(\zeta_{B_2})
}
{(\zeta_{A_1}+\zeta_{B_2})(\zeta_{A_2}+\zeta_{B_1})(\zeta_{A_2}+\zeta_{B_2})}
.
\nonumber \\
\label{G_a}
\eeqy
Here $a_i$ represent unknown weight constants, which will be determined
later, and
$w_{AAB}$ and $w_{BBA}$ represent the continuum counterparts of the amplitudes
$W_{AAB}$ and $W_{BBA}$ respectively.
Similarly the case (b) is classified into two sub cases, (b-1) and (b-2)
(see Fig.~\ref{b}).
The corresponding amplitude is 
\beqy
{\cal G}_b
&=&
b_1\,
\frac{ w_{AAB}(\zeta_{A_1},\zeta_{A_2},\zeta_{B_1})\, w(\zeta_{B_2 }) }
{\zeta_{A_1}+\zeta_{B_2}}
+b_2\,
\frac{ w_{AAB}(\zeta_{A_1},\zeta_{A_2},\zeta_{B_1})\, w(\zeta_{A_1 }) }
{\zeta_{A_1}+\zeta_{B_2}}
\; ,
\eeqy
where $b_i$ are unknown weight constants.
The cases (c) and (d) are obtained by interchanging $A_i$ with $B_i$
 in the cases (b) and (a) respectively:
\beq
{\cal G}_c = {\cal G}_b |_{A\leftrightarrow B}
\; , \quad
{\cal G}_d = {\cal G}_a |_{A\leftrightarrow B}
\; .
\eeq
In the cases (b) and (c), we have doubly counted several terms.
The next amplitude corresponds to the doubly counted terms,
\beqy
{\cal G}_e
&=&
-b_1\,
\frac{ w(\zeta_{A_2})\, w(\zeta_{B_2})
\left(\, w(\zeta_{A_1})+w(\zeta_{B_1}) \, \right) }
{(\zeta_{A_1}+\zeta_{B_1})(\zeta_{A_1}+\zeta_{B_2})(\zeta_{A_2}+\zeta_{B_1})}
\,-b_2\,
\frac{ w(\zeta_{A_1})\, w(\zeta_{B_1})
\left(\, w(\zeta_{A_1})+w(\zeta_{B_1}) \, \right) }
{(\zeta_{A_1}+\zeta_{B_1})(\zeta_{A_1}+\zeta_{B_2})(\zeta_{A_2}+\zeta_{B_1})}
\nonumber \\
&+&
\frac{ w(\zeta_{B_1})\, w(\zeta_{B_2})
\left(\, e_1\, w(\zeta_{A_1})\,+e_2\,w(\zeta_{B_1}) \, \right) }
{(\zeta_{A_1}+\zeta_{B_1})(\zeta_{A_1}+\zeta_{B_2})(\zeta_{A_2}+\zeta_{B_1})}
+\frac{ w(\zeta_{A_1})\, w(\zeta_{A_2})
\left(\, e_1\, w(\zeta_{B_1})\,+e_2\,w(\zeta_{A_1}) \, \right) }
{(\zeta_{A_1}+\zeta_{B_1})(\zeta_{A_1}+\zeta_{B_2})(\zeta_{A_2}+\zeta_{B_1})}
\;.
\nonumber \\
\eeqy
Here the first term comes from the double counting between 
the cases (b-1) and (c-1) (see Fig.~\ref{e1}).
The second term comes from that between 
 the cases (b-2) and (c-2) (see Fig.~\ref{e2}).
The third term represents that between
 the cases (b-1) and (c-2) (see Fig.~\ref{e12}).
The last term represents that between
 the cases (b-2) and (c-1) (see Fig.~\ref{e21}).

The amplitude we aim at is obtained by summing up all the 
cases
\beq
{\cal G}={\cal G}_a +{\cal G}_b+{\cal G}_c+{\cal G}_d+{\cal G}_e
\;.
\label{cG} \eeq
By using the relations
\beq
w_{AAB}(\zeta_1,\zeta_2,\xi)=w_{BBA}(\zeta_1,\zeta_2,\xi)=
-\frac{ w_{AB}(\zeta_1,\xi) -w_{AB}(\zeta_2,\xi) }
{\zeta_1-\zeta_2}
\; ,
\eeq
\beq
w_{AB}(\zeta,\xi)=
\frac{ w(\zeta)^2 +w(\xi)^2 +\,w(\zeta)\,w(\xi)  }
{\zeta+\xi}
\; ,
\label{w_AB_2}
\eeq
we can express $w_{ABAB}$ in terms of the homogeneous disk amplitudes.
It contains 10 unknown weight constants.
They can be decided by requiring symmetries of the amplitude.
Eq.~(\ref{cG}) already has the symmetry under interchanging
$A_i$ with $B_i$.
We further require the symmetry under 
 $(A_1 \leftrightarrow A_2)$ and that under
 $(B_1 \leftrightarrow B_2)$ .
%\hspace*{10mm} $\cdot$ $B_1 \leftrightarrow B_2$ . \\
%\hspace*{10mm} $\cdot$ $A_1 \leftrightarrow A_2$ \\
%\hspace*{10mm} $\cdot$ $B_1 \leftrightarrow B_2$ . \\
The unknown constants are determined up to overall normalization.
The weights $(a_1,a_2,a_4,a_6,b_2)$ have the same non-zero value and
the remaining constants equal to zero.

In this way, $w_{ABAB}$ is determined up to the normalization constant.
When we choose the normalization constant properly ($a_1=1$) ,
the obtained amplitude indeed coincides with that obtained
from the Schwinger-Dyson equations in section 2 \footnote{
   In a similar manner we can also reproduce the terms proportional
   to $t^{4/3}$ in Eq.~(\ref{w_ABAB_fromSD}).}.

%%%%%%%%%%%%%%%%%%%%%%%%%%%%%%%%%%%%%%%%%%%%%%%%%%%%%%%%%%%%%%%
%%%%%%%% section 4 %%%%%%%%%%%%%%%%%%%%%%%%%%%%%%%%%%%%%%%%%%%%
\section{Discussion}

\indspace
In the previous section we derived the heterogeneous disk
amplitude $w_{ABAB}$ only through geometrical consideration.
We required the symmetries of the amplitude there,
however, we consider that the rules of splitting and the consistency
with $w_{AB}$ were essential.

Let us discuss to what extent we can generalize the result we obtained.
According to the rules of loop splitting, we can expect that $w_{AB}$ 
consists of three terms\footnote{
Here we neglect the term proportional to $t^{4/3}$.
}(see Fig.~\ref{AB}).
The rules, however, do not give the information 
on the relative weight of each term.
That is, the weight of the last term of the numerator
of Eq.~(\ref{w_AB_2}) relative to the first two terms
is unknown.
When we treat this weight as an additional unknown
constant, we can also determine it 
in the same way to section 3.
This shows that we can also derive $w_{AB}$ from
the rules of loop splitting and the consistency 
between $w_{AB}$ and $w_{ABAB}$. From this fact, 
we speculate that any heterogeneous loop amplitudes
can be determined from the loop-splitting rules and 
the consistency among the relevant amplitudes.

In boundary conformal field theory, the boundary operator $\sigma$
plays the role to make discontinuity in matter state 
on the boundary.
As we saw in section 3, when it is dressed with gravity, 
it also makes a discontinuity in 
the geometrical state of the boundary.
Here let us comment on the relation with the boundary operators
\cite{MMS,ANA}
which appear in the scaling operators in the matrix models. 
In Ref.~\cite{ANA} it is argued that the scaling operator
 $\hat{\sigma}_{n(m+1)}=\hat{{\cal B}}_n$ has the effect to
 split a loop into $k$ loops in the case of $(m,m+1)$ model,
  where $k\le n$.
When two $\sigma$ approach each other,  one can consider that they
  change the spin state on the boundary locally, and they have the 
 effect to split a loop into two loops \cite{IIKMNS,AI00}.
%The boundary operator $\hat{{\cal B}}_2$ 
%can be identified with the operator which is obtained by
%approaching two $\sigma$ each other.
%In general, we expect that $\hat{{\cal B}}_n$ is obtained by
%approaching $2(k-1)$ $\sigma$ each other, where  $1\le k\le n$.
%
The boundary operator $\hat{B}_2$ can be identified with
the operator obtained in the limit where two $\sigma$
approach each other.
We expect that $\hat{B}_n$ can be constructed by making 
a linear combination of the operators
obtained when $2(k-1)$ spin operators $\sigma$ 
get together, where $1 \le k \le n$.

%%%%%%%%%%%%%%%%%%%%%%%%%%%%%%%%%%%%%%%%%%%%%%%%%%%%%%%%%%%%%%%%%%%%%%%%%%%%%%%
%%%%%%%%%%%%%%%%%%%%%%%%%%% Acknowledgement %%%%%%%%%%%%%%%%%%%%%%%%%%%%%%%%%%%%%
%%%%%%%%%%%%%%%%%%%%%%%%%%%%%%%%%%%%%%%%%%%%%%%%%%%%%%%%%%%%%%%%%%%%%%%%%%%%%%%
\section*{Acknowledgements}

\indspace
We would like to express our gratitude to Professor I. K. Kostov,
because this research has started from the discussion with him.
We are also grateful to Professor M. Ninomiya for warmhearted encouragement.
Thanks are also due to members of YITP, where one of the authors (A.I.)
stayed several times
during the completion of this work.
%\\

%%%%%%%%%%%%%%%%%%%%%%%%%%%%%%%%%%%%%%%%%%%%%%%%%%%%%%%%%%%%%%%%%%%%%%%%%%%%%%
%%%%%%%%%%%%%%%%%%%%%%%%%%%%%  Appendix  %%%%%%%%%%%%%%%%%%%%%%%%%%%%%%%%%%%%%
%%%%%%%%%%%%%%%%%%%%%%%%%%%%%%%%%%%%%%%%%%%%%%%%%%%%%%%%%%%%%%%%%%%%%%%%%%%%%%
\appendix
\section{Inverse Laplace transformation}

\indspace
We summarize here some useful relations on inverse Laplace
transformation.
The Laplace transformation of a function $f(\ell_i)$ is
defined by the relation
\beq
{\cal L}_{i} \bigl[  f(\ell_i) \bigr]
=F(\zeta_i)=
\int_0^\infty d \ell_i \, e^{-\ell_i \zeta_i}\, f(\ell_i)
\;.
\eeq
The following three relations are useful in calculating the 
inverse Laplace transformation,
\beqy
&&{\cal L}^{-1}_{1}{\cal L}^{-1}_{2}
\left[ \frac{1}{\zeta_1 + \zeta_2} \right]
=
\delta(\ell_1 - \ell_2)\,,
\label{formula1}\\
%\eeq
%\beq
&&{\cal L}^{-1}_{1}{\cal L}^{-1}_{2}
\left[ \frac{F(\zeta_1)}{\zeta_1 + \zeta_2} \right]
=
\theta(\ell_1 - \ell_2) \, f(\ell_1 - \ell_2)\,,
\label{formula2}\\
%\eeq
%\beq
&&{\cal L}^{-1}_{i}
\left[ F(\zeta_i)\, G(\zeta_i) \right]
=
\int_0^{\ell_i} d \ell' \, f(\ell')\, g(\ell_i - \ell')
=f(\ell_i)  *_i \,g(\ell_i)
\,.
\label{formula3}
\eeqy
Here $F(\zeta_i)$ and $G(\zeta_j)$ represent the Laplace transformed
functions of $f(\ell_i)$ and $g(\ell_j)$ respectively, and 
the symbol $*_{i}$
in the third equation denotes the convolution with respect to $\ell_i$.
All of the inverse Laplace transformations of the amplitudes
in section 3 can be performed by combining these
relations\footnote{
Eq.~(\ref{formula2}) can be derived combining Eqs.~(\ref{formula1}) 
and (\ref{formula3}).}.

\begin{figure}[htb]
\begin{center}
 \begin{minipage}[htb]{0.33\textwidth}
   \epsfxsize = 1.0\textwidth
   \epsfbox{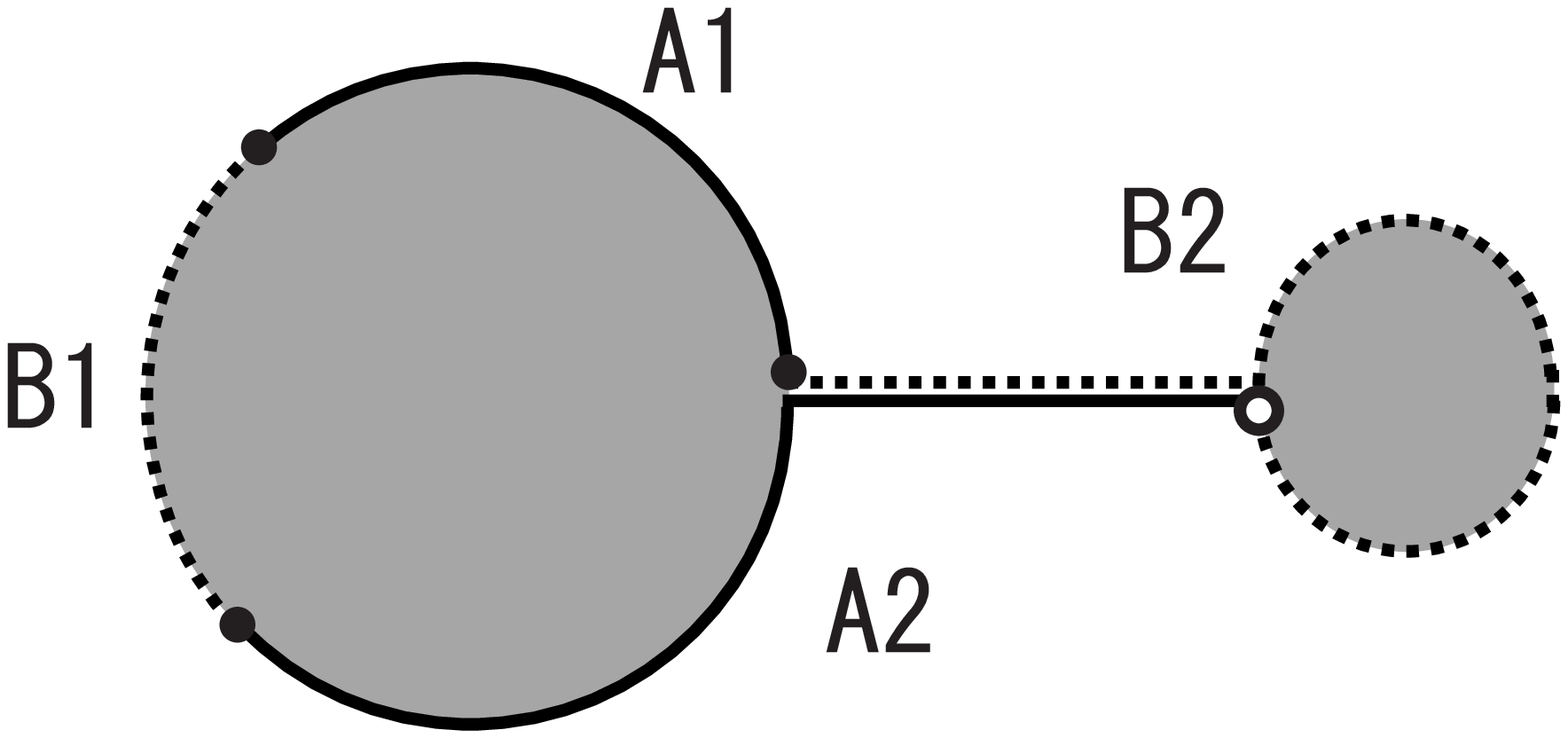}
   \begin{center}\makebox{(a-1)}\end{center}
%   \caption{(a-1)}
%   \label{}
 \end{minipage}
% \hfill
 \hspace{0.4cm}
 \begin{minipage}[htb]{0.33\textwidth}
   \epsfxsize = 1.0\textwidth
   \epsfbox{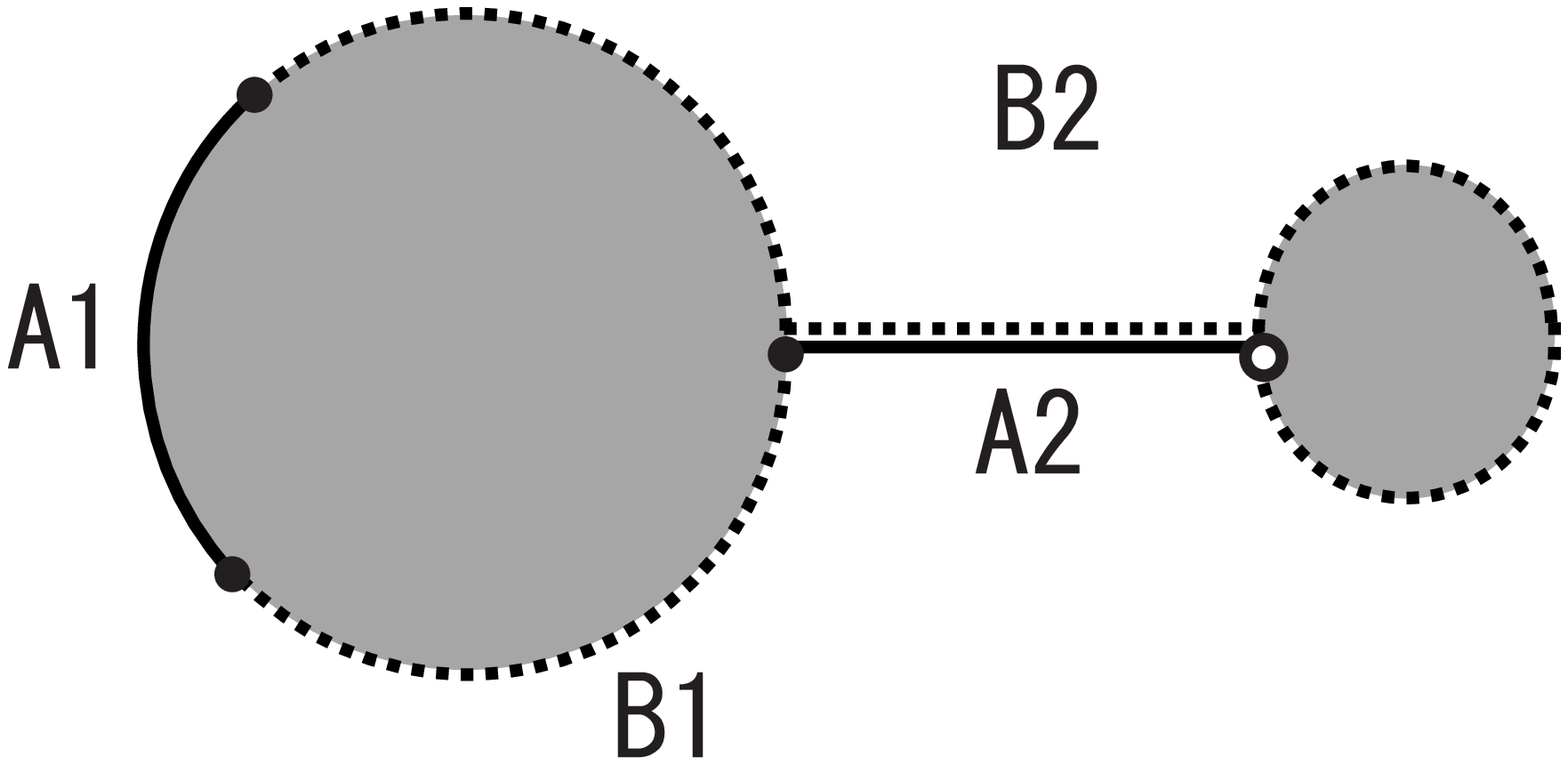}
   \begin{center}\makebox{(a-2)}\end{center}
%  \caption{(a-2)}
%  \label{}
 \end{minipage}
  \hspace{0.4cm}
 \begin{minipage}[htb]{0.26\textwidth}
   \epsfxsize = 1.0\textwidth
   \epsfbox{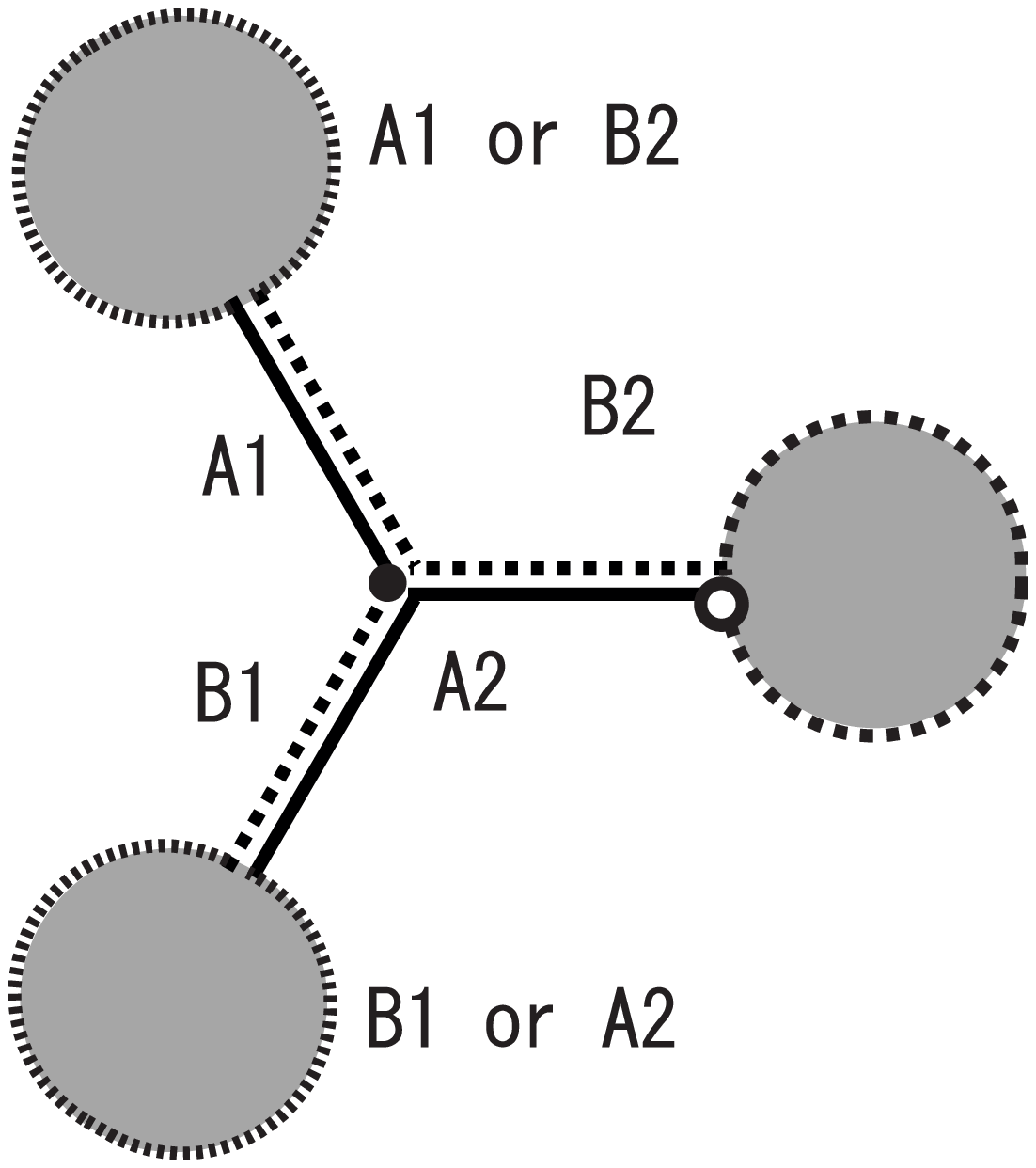}
   \begin{center}\makebox{(a-3)}\end{center}
%  \caption{(a-3)}
%  \label{}
 \end{minipage}
 \caption{
   The case (a) is furthermore classified into the three sub cases.
 }
 \label{a}
\end{center} 
\end{figure}

\begin{figure}[ht]
\begin{center}
 \begin{minipage}[htb]{0.26\textwidth}
   \epsfxsize = 1.0\textwidth
   \epsfbox{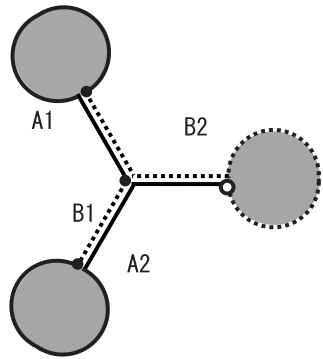}
   \caption{
     The diagram corresponding to Eq.~(\ref{example2})
   }
   \label{app_a3}
 \end{minipage}
% \caption{}
% \label{}
\end{center} 
\end{figure}

As examples, we show the inverse Laplace transformations
of some terms in Eq.~(\ref{G_a}) in the following:
\beqy
\lefteqn{
{\cal L}^{-1}_{A_1} {\cal L}^{-1}_{A_2}
{\cal L}^{-1}_{B_1} {\cal L}^{-1}_{B_2}
\left[
\frac{ w_{AAB}(\zeta_{A_1},\zeta_{A_2},\zeta_{B_1})\, w(\zeta_{B_2 }) }
{\zeta_{A_2}+\zeta_{B_2}}
\right]
} \nonumber \\
&&\quad=
\bigl[ \theta(\ell_{A_2}-\ell_{B_2})
 \cW_{AAB}(\ell_{A_1}, \ell_{A_2}-\ell_{B_2}, \ell_{B_1}) \bigr]
*_{B_2} \, \cW(\ell_{B_2})
\nonumber \\
&&\quad=
\int_0^{{\rm min}(\ell_{A_2}, \ell_{B_2})} \!\!\! d\ell' \;
 \cW_{AAB}(\ell_{A_1}, \ell_{A_2}-\ell', \ell_{B_1})
 \, \cW(\ell_{B_2}-\ell')
\, ,
\label{example1}
\eeqy
\beqy
\lefteqn{
{\cal L}^{-1}_{A_1} {\cal L}^{-1}_{A_2}
{\cal L}^{-1}_{B_1} {\cal L}^{-1}_{B_2}
\left[
\frac{ w(\zeta_{A_1})\, w(\zeta_{A_2})\, w(\zeta_{B_2 }) }
{(\zeta_{A_1}+\zeta_{B_2})(\zeta_{A_2}+\zeta_{B_1})
 (\zeta_{A_2}+\zeta_{B_2})}
\right]
} \nonumber \\
&&\quad=
\bigl[ \theta(\ell_{A_1}-\ell_{B_2}) \cW(\ell_{A_1}-\ell_{B_2})
\bigr]
 \nonumber \\
&&\qquad
*_{B_2} \,
\Bigl\{
\bigl[ \theta(\ell_{A_2}-\ell_{B_1}) \cW(\ell_{A_2}-\ell_{B_1})
\bigr]
*_{A_2} \,
\bigl[ \theta(\ell_{B_2}-\ell_{A_2}) \cW(\ell_{B_2}-\ell_{A_2})
\bigr]
\Bigr\}
\, .
\label{example2}
\eeqy From Eq.~(\ref{example1}), 
one finds that this term corresponds to 
the first diagram in Fig.~\ref{a}.
Similarly  the amplitude (\ref{example2})
 can be represented by Fig.~\ref{app_a3}.

\newpage
%%%%%%%%%%%%%%%%%%%%%%%%%%%%%%%%%%%%%%%%%%%%%%%%%%%%%%%%%%%%%%%%%%%%%%%%%%%%%%%
%%%%%%%%%%%%%%%%%%%%%%%%%%%%% References %%%%%%%%%%%%%%%%%%%%%%%%%%%%%%%%%%%%%%
%%%%%%%%%%%%%%%%%%%%%%%%%%%%%%%%%%%%%%%%%%%%%%%%%%%%%%%%%%%%%%%%%%%%%%%%%%%%%%%

\newpage
%%%%%%%%%%%%%%%%%%%%%%%%%%%%%%%%%%
%%%%%%%%%%%%% FIGURES %%%%%%%%%%%%
%%%%%%%%%%%%%%%%%%%%%%%%%%%%%%%%%%
%%%%%%%%%%%%%%%%%%%%%%%%%%%%%%%%%%%%%
%%%%%%%% 1st Line (1 picture)%%%%%%%%
%%%%%%%%%%%%%%%%%%%%%%%%%%%%%%%%%%%%%
%%%%%%%%%% End 1st Line %%%%%%%%%%%%%
%\vspace{10mm}
%%%%%%%%%%%%%%%%%%%%%%%%%%%%%%%%%%%%%
%%%%%%%% 2nd Line (3 pictures)%%%%%%%
%%%%%%%%%%%%%%%%%%%%%%%%%%%%%%%%%%%%%
%%%%%%%%%% End 2nd Line %%%%%%%%%%%%%
%\vspace{10mm}
%%%%%%%%%%%%%%%%%%%%%%%%%%%%%%%%%%%%%
%%%%%%%% 3rd Line (2 picture)%%%%%%%%
%%%%%%%%%%%%%%%%%%%%%%%%%%%%%%%%%%%%%
%%%%%%%%%% End 3rd Line %%%%%%%%%%%%%
%\vspace{10mm}
%%%%%%%%%%%%%%%%%%%%%%%%%%%%%%%%%%%%%
%%%%%%%% 4th Line (3 pictures)%%%%%%%
%%%%%%%%%%%%%%%%%%%%%%%%%%%%%%%%%%%%%
%%%%%%%%%% End 4th Line %%%%%%%%%%%%%
%\vspace{10mm}
%%%%%%%%%%%%%%%%%%%%%%%%%%%%%%%%%%%%%
%%%%%%%% 5th Line (2 pictures)%%%%%%%
%%%%%%%%%%%%%%%%%%%%%%%%%%%%%%%%%%%%%
%%%%%%%%%% End 5th Line %%%%%%%%%%%%%
\vspace{5mm}
%%%%%%%%%%%%%%%%%%%%%%%%%%%%%%%%%%%%%
%%%%%%%% 6th Line (2 pictures)%%%%%%%
%%%%%%%%%%%%%%%%%%%%%%%%%%%%%%%%%%%%%
%%%%%%%%%% End 6th Line %%%%%%%%%
\vspace{5mm}
%%%%%%%%%%%%%%%%%%%%%%%%%%%%%%%%%%%%%
%%%%%%%% 7th Line (2 pictures)%%%%%%%
%%%%%%%%%%%%%%%%%%%%%%%%%%%%%%%%%%%%%
%%%%%%%%%% End 7th Line %%%%%%%%%
\vspace{5mm}
%%%%%%%%%%%%%%%%%%%%%%%%%%%%%%%%%%%%%
%%%%%%%% 8th Line (1 picture)%%%%%%%
%%%%%%%%%%%%%%%%%%%%%%%%%%%%%%%%%%%%%
%%%%%%%%%% End 7th Line %%%%%%%%%

\end{document}